# Polarization observations of SNR G156.2+5.7 at λ6cm

J. W. Xu[1], J. L. Han[1], X. H. Sun[1,2], W. Reich[2], L. Xiao[1], P. Reich[2], and R. Wielebinski[2]

[1] National Astronomical Observatories, Chinese Academy of Sciences, Beijing 100012, China
   e-mail: xjw, hjl, xl, xhsun@bao.ac.cn

[2] Max-Planck-Institut für Radioastronomie, Auf dem Hügel 69, 53121 Bonn, Germany
   e-mail: wreich, xhsun, preich, rwielebinski@mpifr-bonn.mpg.de

**ABSTRACT**

*Context.* G156.2+5.7 is a large supernova remnant (SNR) first discovered in the ROSAT X-ray survey due to its exceptional high X-ray brightness. Subsequently detailed X-ray, optical and Hα studies of this SNR were made, but radio observations are rare because of its low surface brightness and large size.

*Aims.* Radio continuum and polarization images of the SNR G156.2+5.7 at λ6 cm are presented for the first time to be discussed in the context with available data to investigate its physical properties.

*Methods.* The SNR G156.2+5.7 was observed with the Urumqi 25 m telescope in radio continuum and linear polarization. The results are analyzed together with 11 cm and 21 cm maps previously obtained with the Effelsberg 100 m telescope and compared with a broad-band X-ray ROSAT image, 100 μm infrared radiation data and a Hα emission map.

*Results.* We obtained an integrated flux density for G156.2+5.7 of $S_{6cm} = 2.5 \pm 0.5$ Jy. The spectral index of the integrated emission is $\alpha = -0.48 \pm 0.08$ (S~ $\nu^{\alpha}$) between λ74 cm and λ6 cm excluding compact radio sources. We also obtained a spectral index map, which shows little variation between the shells and the central area of the SNR. This is consistent with the constant integrated flux density spectrum. Highly polarized radio emission has been detected from the SNR shell, but also from a central patch, which probably originates in the front part of the spherical SNR shell. We derived the distribution of rotation measure from polarization data at λ11 cm and λ6 cm and found RM gradients of opposite direction in the SNR shell.

*Conclusions.* The SNR G156.2+5.7 is unusual by its exceptionally high X-ray brightness and very low surface-brightness in the radio range. The magnetic field is very well ordered along the shell periphery as expected for a compressed ambient magnetic field. A toroidal magnetic field component is indicated by the RM distribution.

## 1. Introduction

The SNR G156.2+5.7 is the first Galactic supernova remnant discovered by its X-ray emission (Pfeffermann et al. 1991). In the discovery image taken from the X-ray all-sky survey by ROSAT SNR G156.2+5.7 appears to be an almost circular bright object of 108′ in diameter. Subsequent follow-up radio observations at λ21 cm and λ11 cm with the Effelsberg 100 m telescope (Reich et al. 1992) show a clear radio shell structure with highly polarized rims, whose characteristics confirm the identification of G156.2+5.7 as a SNR. Reich et al. (1992) noticed that the radio surface brightness of G156.2+5.7 is only $5.8 \times 10^{-23}$ Wm$^{-2}$Hz$^{-1}$sr$^{-1}$ at 1 GHz, one of the lowest among all known SNRs (see Xu et al. 2005; Green 2005). Very interesting results come from recent optical Hα images and spectral line observations of the filaments (Gerardy & Fesen 2007), which show that some Hα filaments are produced by the supernova shock and associated with the boundary of the central hot gas bubble as revealed by the X-ray emission.

All authors agreed that the SNR G156.2+5.7 is likely in the phase of adiabatic expansion (i.e. Sedov phase), which is based on its regular shell-type morphology and its radio spectral index. However, the basic parameters of SNR G156.2+5.7 are still not well settled. Assuming an initial kinetic energy of $E_0 = 10^{51}$ erg, Pfeffermann et al. (1991) estimated a shock velocity of $v_s \sim 651\,\mathrm{km\,s^{-1}}$, a distance of $d \sim 3$ kpc, a diameter of $D \sim 95$ pc, an age of $\tau \sim 2.6 \times 10^4$ yr, $n_0 \sim 0.01$ cm$^{-3}$ and $L_x \sim 2.1\,10^{35}$ erg s$^{-1}$ (0.1 - 2.4 keV), and the mass swept up during the expansion $M_{sw} \sim 142 M_\odot$. Reich et al. (1992) analyzed HI-data for associated emission and derived a distance between 3 kpc and 1 kpc, meaning that the diameter $D$ of the radio shell is between 100 pc and 32 pc. The HI-shell, which was most likely created by the stellar wind of the progenitor star is larger than the radio shell, and the swept-mass was calculated to be 6300 $M_\odot$ or 700 $M_\odot$ for a distance of 3 kpc or 1 kpc, respectively. After considering non-equilibrium ionization effects in the low-density plasma, i.e. the ionization equilibrium time can be large then the age of the SNR, Yamauchi et al.

**Table 1.** Parameters of Urumqi 6cm system and scan observations

| | |
|---|---|
| Frequency (GHz) | 4.8 |
| Bandwidth (MHz) | 600 |
| $T_{sys}$ (K) | 22 |
| $T_B[K]/S[Jy]$ | 0.164 |
| aperture efficiency | 62% |
| Beam size HPBW (') | 9.5 |
| Scan directions | $l$ & $b$ |
| Scan velocity [$^o$/min] | 2 |
| Sub-scan separation ['] | 4 |

(1999) fitted the ASCA X-ray data in the north (left in Galactic coordinates) rim and central region, and they estimated the age of $\tau \sim 1.5 \times 10^4$ yr, a distance of $d \sim 1.3$ kpc, $n_0 \sim 0.2$ cm$^{-3}$, and $M_{sw} \sim 110 M_\odot$. The diameter then should be 40 pc. The optical images and the line ratios of optical spectra suggest that the remnant is probably interacting with a clumpy interstellar medium (Gerardy & Fesen 2007). For a pre-shock density of 10 cm$^{-3}$ the distance of the SNR is between 0.3 kpc and 0.6 kpc. In this case the radius of the remnant is between 5 pc and 10 pc, and the age is between 400 yr and 2200 yr. The shock velocity ranges between $\sim 5000$ km s$^{-1}$ or 1500 km s$^{-1}$. There is consensus that the location of the SNR is between the local arm and the Perseus arm. The Perseus arm's distance is about 2 kpc at the Galactic longitude $GL \sim 134°$ (Xu et al. 2006), so that we adopt a distance of 1 kpc in the following and assume the diameter of SNR G156.2+5.7 to be 32 pc.

As mentioned above, SNR G156.2+5.7 has not been mapped in many bands, because it is too large and too faint for many telescopes. We observed this large and faint SNR with the $\lambda 6$ cm system installed at the Urumqi 25 m telescope in total intensity and linear polarization. In this paper we present a thorough analysis of available radio data and compare them with published images obtained in other bands. The observations of SNR G156.2+5.7 at $\lambda 6$ cm and results are briefly described in Sect. 2. In Sect. 3 we discuss its spectral and polarization properties. We compare the radio morphology with images from other bands and discuss the physical properties of G156.2+5.7 in Sect. 4. We summarize our results finally in Sect. 5.

## 2. Observations and radio data at $\lambda 6$ cm

During the test phase of observations for the Sino-German 6 cm polarization survey of the Galactic plane, we observed the SNR G156.2+5.7 with the 25 m telescope at Nanshan station of the Urumqi Observatory of National Astronomical Observatories, Chinese Academy of Sciecences (NAOC). The 6 cm continuum and polarization observation system mounted at the telescope was constructed at the Max-Planck-Institut für Radioastronomie (MPIfR)/Germany, as being a special support to the Partner Group of MPIfR at NAOC. The system has been already described by Sun et al. (2006, 2007). The basic observational parameters are listed in Table 1.

The observations were carried out in scanning mode either along Galactic longitude or latitude. In total 8 maps were obtained. The raw data consist of the signals from the four backend channels (RR*, LL*, RL*, LR*) modulated every 16 msec by a 1.7 K calibration signal. The scan data together with time and telescope position information were stored on a Linux PC.

The data were converted into the NOD2 format (Haslam 1974) for further processing. The data reduction package was adopted from the MPIfR and converted from the Unix to the LINUX operation system. Spiky interferences in each of the four channel maps of every coverage were removed and any base-line curvature of each sub-scan was corrected by an appropriate polynomial fit. Scanning effects were suppressed by applying an "unsharp masking method" (Sofue & Reich 1979). The individually processed data from all coverages were combined in the Fourier plane using the method as described by Emerson & Gräve (1988), which includes a further reducing of residual scanning effects left from the "unsharp masking method", which uses low order polynomial fits for correction. The final maps from the four channels were combined into maps of the Stokes parameter $I$, $Q$ and $U$. For an increase of the S/N ratio we slightly smoothed the final $\lambda 6$ cm maps to a circular Gaussian beam size of 10'. The polarized intensity ($PI$) map was calculated from the $U$ and $Q$ maps taking their r.m.s.– noise of $\sigma_{U,Q} = 0.3$ mK $T_B$ into account by $PI^2 = U^2 + Q^2$ - $1.2\sigma^2$. The polarization angle ($PA$) map was calculated from the $U$ and $Q$ maps by $PA = 0.5$ atan$(U/Q)$.

We observed the sources 3C138 and 3C48 for the calibration of the flux density scale and the polarization data. The flux density of 3C138 and 3C48 were assumed to be 3.9 Jy and 5.5 Jy at 6 cm respectively. A percentage polarization of 10.8% and 4.2% was assumed for 3C138 and 3C48, respectively. The polarization angle of 3C138 was taken to be 169°.

The radio image of SNR G156.2+5.7 observed at $\lambda 6$ cm is shown in Fig. 1 being slightly convolved to 10'. The r.m.s.– noise was measured to be about 0.75 mK $T_B$ in the total intensity map. The total intensity noise exceeds that measured from the cross-correlated U and Q data, which are less depended on receiver gain fluctuations and the influence of weather effects on short time scales.

There are a number of point-like radio sources in the observed field of the SNR, which are considered as extragalactic background sources. The strongest point-like source located in the southwest of the remnant ($l = 155°.361$, $b = 5°.116$) is the radio galaxy 3C130, which was already discussed by Reich et al. (1992).

After subtracting the contribution from compact point sources, we obtained an integrated flux density of the SNR of 2.5±0.5 Jy in total intensity. The shell structure is more pronounced in polarized intensity than in total intensity. The magnetic field lines roughly follow the SNR-shell, as indicated by plotting the polarization E-vectors by $PA + 90°$. This assumes that Faraday rotation is not too large (see Sect. 3.2). A remarkable feature is the well defined central patch in the polarization map.

## 3. Radio properties of the SNR G156.2+5.7

In this section, we investigate the radio properties, including the spectral distribution, the polarization distribution as well as the distribution of the rotation measure of SNR G156.2+5.7 using radio images at $\lambda\lambda 6$ cm, 11 cm and 21 cm. We used the Effelsberg maps observed at λ11 cm (Fig. 2) and λ21 cm (Fig. 3) from Reich et al. (1992). In general the main features of the three images are rather similar in total intensity. Also the radio image at 408 MHz published by Kothes et al. (2006) in their Fig. 6 shows the same properties.

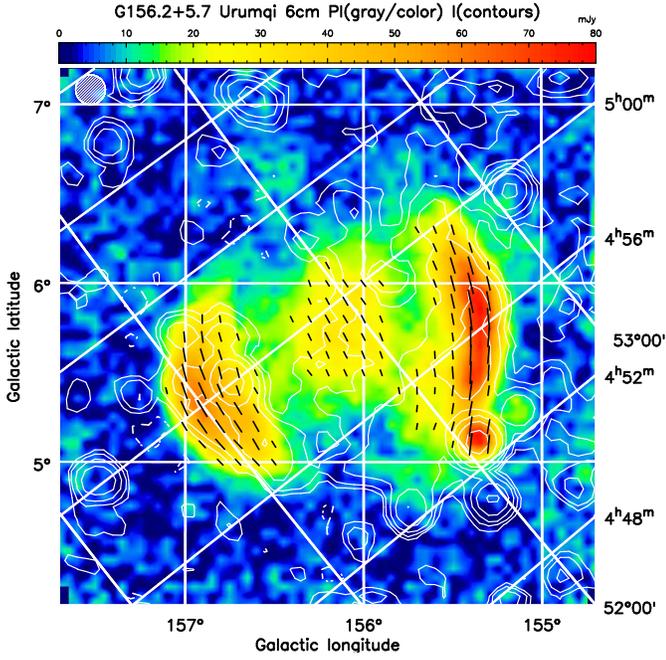

**Fig. 1.** Radio image of SNR G156.2+5.7 at λ6 cm. Contours are shown for total intensities at $\pm 2^n \sigma$ (n = 1, 2, 3, 4, ...). Negative contours are indicated by dash lines. Here $\sigma$ = 0.75 mK or 5.0 mJy/beam area. Polarized intensities are displayed in gray scale with a $\sigma$ of about 0.3 mK or 2.0 mJy/beam area. A central polarized patch is clearly visible. The magnetic field directions ($PA + 90°$ for small Faraday rotation) are indicated by vectors, whose lengths are proportional to the polarized intensity. The beam size (10′) is marked in the upper left corner. The coordinate grids of right ascention and declination (J2000) are also shown in this map for easy comparison with images in literature.

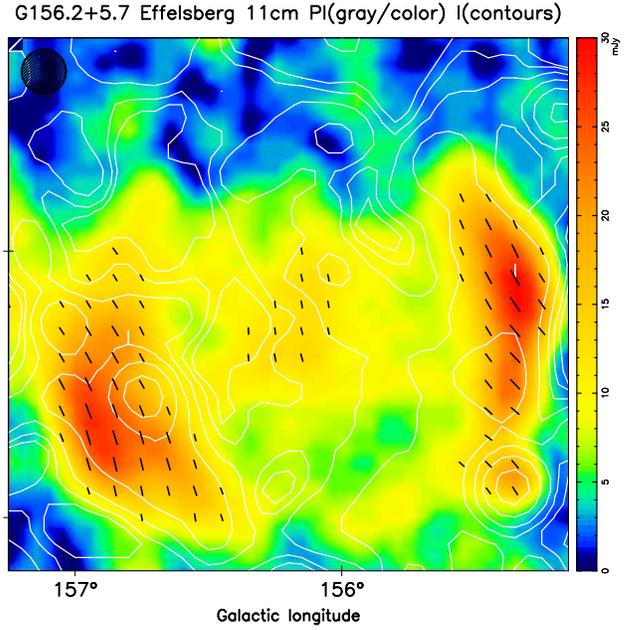

**Fig. 2.** Radio image of SNR G156.2+5.7 at λ11 cm displaying the same quantities as in Fig. 1. The $\sigma$ of total intensities and polarized intensities is 1.5 mJy/beam area and 0.8 mJy/beam area, respectively. The beam size is 10′. Data were taken from Reich et al. (1992).

### 3.1. Spectral index analysis

The T-T plot method has been widely used for the calculation of spectral indices of extended radio sources. This method largely reduces the influence of unrelated large-scale background emission on the spectral index for a selected object. After convolving the maps obtained at two frequencies to the same resolution, the values of map pixels for a selected area are plotted against each other. A linear regression gives the relative background offset and the slope, $\beta$, is the average spectral index for the selected object. $\beta$ is the brightness temperature spectral index for the source ($T_1$ and $T_2$) in the region, i.e. ($T \propto \nu^\beta$). The flux density spectral index, $\alpha$, defined as $S \propto \nu^\alpha$, relates to $\beta$ by $\beta = \alpha - 2$.

The T-T plots for the entire SNR G156.2+5.7 ($2° \times 2°$) have been obtained based on the three available maps, e.g. the Urumqi λ6 cm and the Effelsberg λ11 cm and λ21 cm maps after being smoothed to the same angular resolution of 10′ (see Fig. 4 and Tab. 2). We have discounted the effect of com-

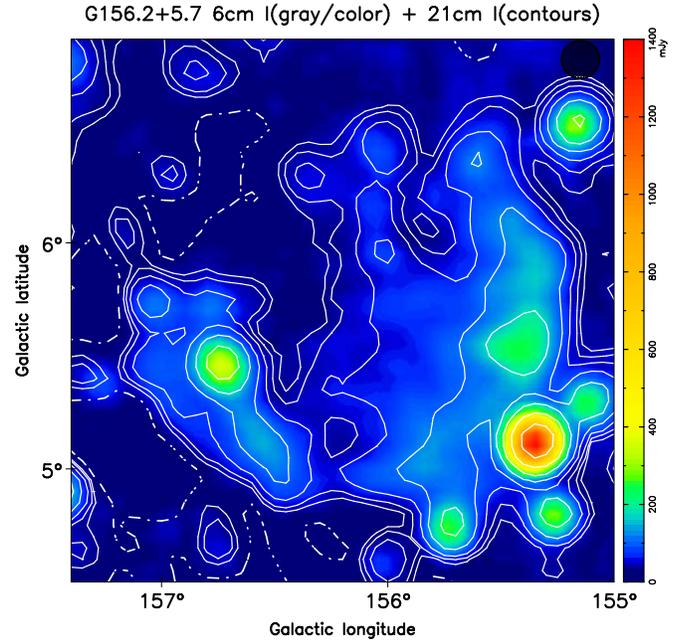

**Fig. 3.** Contour map of SNR G156.2+5.7 at λ21 cm for the total intensity overlayed onto the λ6 cm total intensity map in the color/gray scale. The levels are -2 (dashed) and $2^n$(n = 1, 2, 3, ...) ×$\sigma$, where the r.m.s.–noise $\sigma$ is 5 mJy/beam area. The common beam size of 10′ of both maps is shown in the lower left corner. Data at λ21 cm were taken from Reich et al. (1992).

pact radio sources with a brightness strong enough to apply a Gaussian fit and we obtained spectral indices $\beta = -2.56 \pm 0.21$ from the λ6 cm/λ11 cm maps and $\beta = -2.54 \pm 0.07$ from λ6 cm/λ21 cm maps. These spectral indices are in good agree-

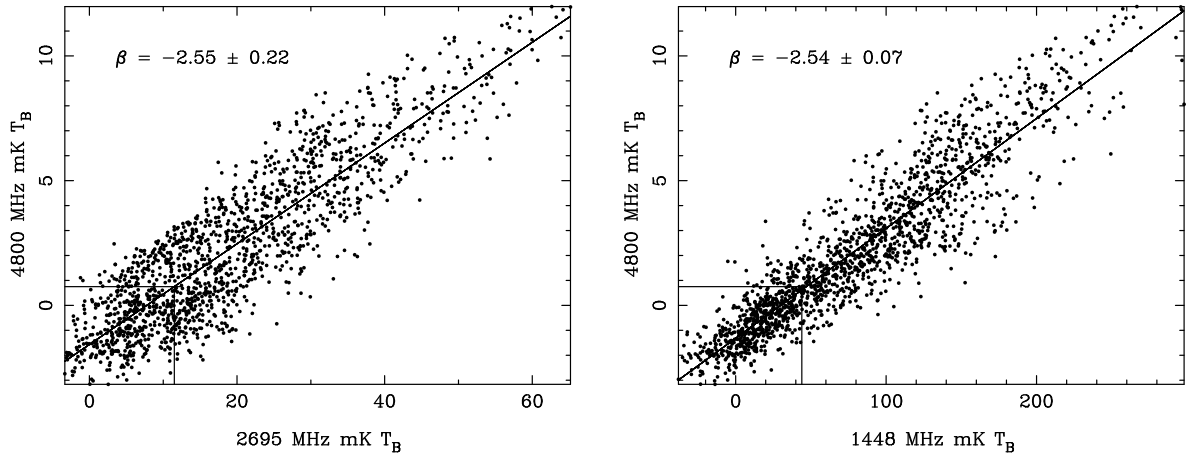

**Fig. 4.** The T-T plots for the entire area of SNR G156.2+5.7 after removing compact radio sources.

**Table 2.** Temperature spectral indices (β) of G156.2+5.7 obtained through TT-plots for three pairs of wavelengths for various areas: Here 'left' and 'right' refer to the images shown in Galactic coordinates and are used throughout the text. In the equatorial coordinate system, the right rim is located to the North, and the left rim to the South.

| Region | $\beta(6/11)$ | $\beta(6/21)$ | $\beta(11/21)$ |
|---|---|---|---|
| Entire remnant | −2.55±0.22 | −2.54±0.07 | −2.54±0.22 |
| Left area | −2.64±0.15 | −2.58±0.05 | −2.50±0.16 |
| Central area | −2.82±0.42 | −2.84±0.08 | −2.86±0.24 |
| Right area | −2.68±0.26 | −2.49±0.09 | −2.35±0.27 |

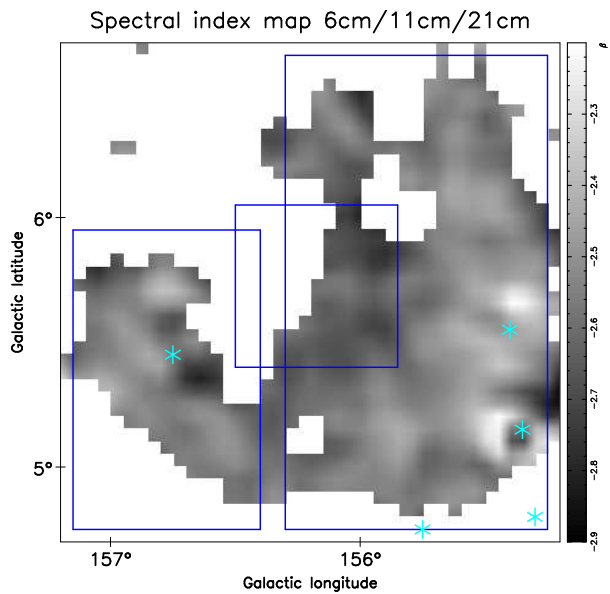

**Fig. 5.** Temperature spectral index map for SNR G156.2+5.7 computed from radio maps at λ6cm, λ11cm and λ21cm. All maps are convolved to an angular resolution of HPBW = 10'. The gray scale wedge ranges from −2.9 (dark) to −2.2 (light). The point sources which were discounted for the calculation are marked by '*', and the regions mentioned in Table 2 are indicated by boxes.

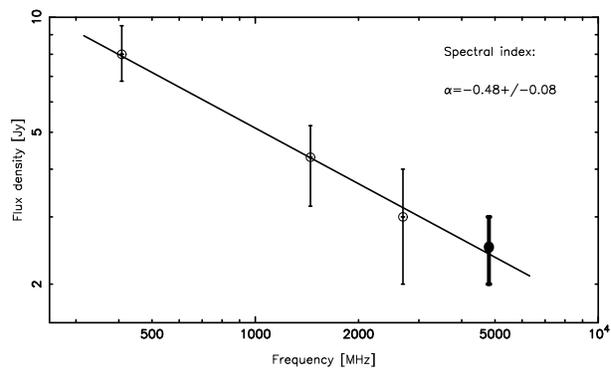

**Fig. 6.** The spectrum for integrated flux densities of the SNR G156.2+5.7. Flux densities from Kothes et al. (2006) and Reich et al. (1992) are shown by open circles. The new λ6 cm measurement is indicated by a dark dot.

To check for possible variations of the spectral index in different SNR regions, we have performed a more detailed analysis. First we obtained spectral indices based on TT-plots for three selected regions, i.e. the left shell, the central region and the right shell of the SNR, using again the three pairs of maps (the results are listed in Tab. 2). The smallest error for the spectral index value was achieved for the λ6cm/λ21cm pair having the largest frequency separation. A second method is to directly calculate a spectral index map (Fig. 5) from the three radio maps with adjusted background levels. Cut-off levels as shown in the TT-plots were set to diminish the influence of systematic effects at low intensity levels and to use data only from areas with a high signal-to-noise ratio. There is an indication both from the T-T plot method and the spectral index map that the central region of the SNR has a slightly steeper spectrum than both shells, which have quite a similar spectral index of about $\alpha = -0.5$.

The flux densities at the λ21cm, 11cm and 6cm have been recalculated after the maps were 'twisted' to the same zero level at the four edge areas and compact point-like sources were subtracted. We got an integrated flux density at λ21cm of 4.3±1.0Jy, slightly larger than that in Reich et al. (1992). The recomputed flux density at λ11cm is 3.0±1.0Jy, exactly ment with those previously derived by Reich et al. (1992), i.e., $\beta = -2.50 \pm 0.15$ from the λ21cm/λ11cm maps.

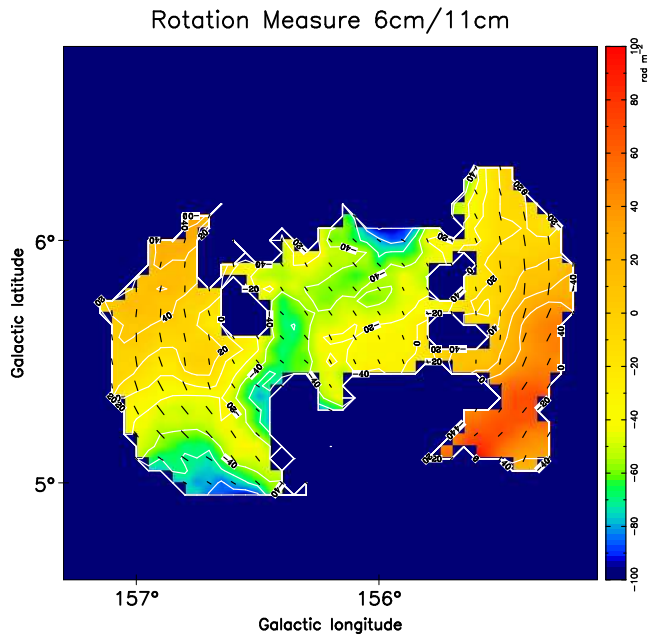

Rotation Measure 6cm/11cm

**Fig. 7.** The RM map of the SNR G156.2+5.7 calculated from the polarization angle maps at λ6 cm and λ11 cm. The $U$ and $Q$ maps at the two frequencies were used at 10′ angular resolution for this purpose. No RMs were calculated for regions where the polarized intensity was less than 3.5 σ at either frequency.

the same value as in Reich et al. (1992). The integrated flux density of G156.2+5.7 at λ6 cm is 2.5±0.5 Jy. The quoted error includes the uncertainties in determining the calibration factors and background levels, which adds up in total to about 2%.

Combining our integrated λ6 cm flux density together with those at lower frequencies, including the flux density at 408 MHz by Kothes et al. (2006), we obtain a spectral index for integrated flux densities of $\alpha = -0.48 \pm 0.08$ (Fig. 6). This is consistent with $\alpha = -0.50\pm0.15$ given by Reich et al. (1992) from data at λ21 cm and λ11 cm. Clearly, there is no sign of a spectral break in the frequency range up to λ6 cm. The average spectral index $\alpha = -0.54 \pm 0.07$ obtained from the TT-plots is just slightly steeper but clearly consistent with that derived from integrated flux densities at three wavelengths.

### 3.2. The distribution of rotation measure

From the polarization observations of the SNR G156.2+5.7 at λ6 cm and λ11 cm we derived the rotation measure (RM) map of this SNR. With only two wavelengths the ambiguity of the calculated RM values needs to be considered, which is ±370 rad m$^{-2}$. Because this SNR is at about 1 kpc distance RM values as large as 370 rad m$^{-2}$ or −370 rad m$^{-2}$ can be most likely ruled out, that the directly derived RMs with the smallest average |RM| value are taken as the most likely RM map of the SNR.

The $U$ and $Q$ maps at the two wavelengths were used at an angular resolution of 10′ to calculate the $PA$ maps and $PI$ maps. RM values were calculated from the two $PA$ maps only when the polarized intensity at any pixel in the $PI$ map at each frequency is greater than $3.5\sigma$. The RM map derived that way is shown in Fig. 7. The largest RM values in the map ranging between +120 rad m$^{-2}$ and +175 rad m$^{-2}$ are seen in the area close to the strong compact source 3C130 (see Sect. 2) and may result from some spurious polarization not related to the SNR G156.2+5.7. For the area of the SNR the RMs vary between +130 rad m$^{-2}$ and -100 rad m$^{-2}$. The mean RM is about +5 rad m$^{-2}$, which is a rather typical RM value for the diffuse Galactic polarized emission at long wavelengths for this region of the Galaxy (Spoelstra 1984). The mean RM for the right shell ridge is about +40 rad m$^{-2}$, for the central patch about -32 rad m$^{-2}$ and -7 rad m$^{-2}$ for the left shell ridge. The dispersion in all areas is between 20 rad m$^{-2}$ and 40 rad m$^{-2}$. Remarkable is a gradient along both ridges running for the right shell ridge from +120 rad m$^{-2}$ close to 3C130 to about +5 rad m$^{-2}$ at its upper end. The gradient along the left shell is in the opposite direction running approximately from about -80 rad m$^{-2}$ to +70 rad m$^{-2}$ from bottom to top. The RM gradients are certainly too strong to be explained by variations of the magnetic field or the electron density in the interstellar medium along the line-of-sight. If the ambient interstellar magnetic field is homogen in the area surrounding G156.2+5.7 we would not expect such RM gradients, but about the same RM values for both shell ridges depending on the inclination of the magnetic field or SNR axis relative to the line-of-sight. Thus the opposite RM gradients observed for the SNR G156.2+5.7 may indicate the existence of a weak toroidal magnetic field component in the SNR shell. Such a RM configuration has not been observed so clearly for other SNRs before. If this toroidal component is swept up from the ambient Galactic magnetic field or results from the evolution of the SNR within the interstellar medium can not be decided from the existing data.

### 3.3. Polarization

Polarized emission is detected from the SNR G156.2+5.7 at λ6 cm mainly from three patches, the left shell ridge, the central area and the right shell ridge (see Fig. 1). A significantly polarized feature is the central patch, well separated from both ridges of the shell. The highest percentage polarization value is 65% for the left shell ridge and the central area and 50% for the right shell. The approximate magnetic field direction in the two shell ridges as indicated by the bars in Fig. 1 follows the shell periphery. In the highly polarized central patch the magnetic field direction is roughly orientated along the symmetry axis of the two shell ridges.

All available data indicate that G156.2+5.7 is an evolved SNR and therefore the magnetic field in the shell should be compressed by the supernova shock. The magnetic field structure of the SNR should reflect the magnetic field in the ambient interstellar medium. The interstellar magnetic field is mostly confined to the Galactic disk and parallel to the Galactic plane (Heiles 1996; Han & Qiao 1994). In such an environment SNRs at low Galactic latitude tend to have a bilateral axis aligned with the Galactic plane (Gaensler 1998), and the magnetic field lines of evolved SNRs should follow the shell and be symmetric to the bilateral axis. The SNR G156.2+5.7 is located at a few degrees latitude outside the Galactic plane in the thin in-terstellar environment of an interarm region towards the anticentre direction of the Galaxy. From the polarization maps of SNR G156.2+5.7 (Fig. 1 and Fig. 2), we can infer that the ambient magnetic fields should have the orientation parallel to the symmetric axis of the SNR polarization angles, i.e. with an angle of about 60° inclined from the Galactic plane. This inclination, however, is exceptional large as seen from the compilation of inclination angles for bilateral SNRs by Gaensler (1998), which indicates that the vertical component of the ambient magnetic field $B_z$ should be about two times stronger than the azimuthal magnetic field $B_\theta$ in this anti-central region at this Galactic latitude.

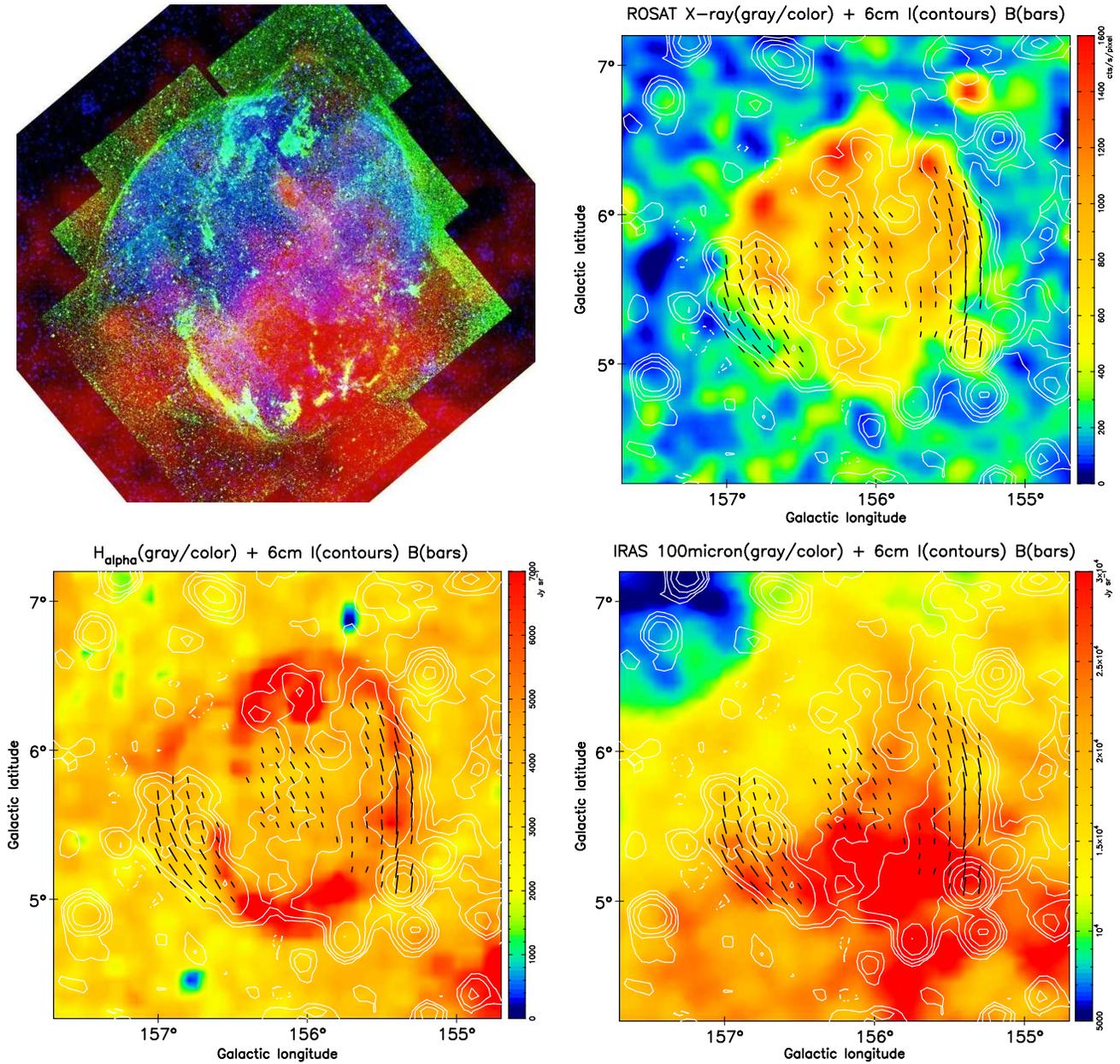

**Fig. 8.** Multiband images for the SNR G156.2+5.7. Upper left: superimposed X-ray image (blue), the recently published high resolution $H\alpha$ image (green) and the $A_v$ extinction (red) of SNR G156.2+5.7. The image is taken from Fig. 9 of Gerardy & Fesen (2007) but rotated by ∼ 50° clockwise to align with the Galactic coordinate system. Upper right: The superposition of a ROSAT X-ray image (Voges et al. 1999) and contours showing $\lambda$6 cm total intensities and vectors showing the magnetic field direction. Lower left: The superposition of a low resolution $H\alpha$ image (Finkbeiner 2003) and with $\lambda$6 cm data as before. Lower right: The superposition of $100\,\mu$m IRAS image (Wheelock et al. 1991) and the $\lambda$6 cm data as before.

Polarization structures at $\lambda$6 cm and $\lambda$11 cm are morphologically very similar (see Fig. 1 and Fig. 2), but the central polarized patch is much less pronounced at 11 cm (Reich et al. 1992). Such a distinct central feature might be taken as an indication for the existence of a central plerionic component inside the SNR shell, but there is no hint for an associated total intensity flat spectrum emission component to support such an interpretation. Therefore we believe that this central polarized structure is an intrinsic structure of the SNR shell, probably from the front part of the shell, where the magnetic field is very well ordered as indicated by its high percentage polarization. Nevertheless we note that such central polarized patches are unusual in shell-type SNRs. The only other case we found in the literature is the SNR G39.2-0.3 (see Fig. 2 of Becker & Helfand 1987).Except for the central polarization patch, polarization properties of G156.2+5.7 are similar to that of another shell-type SNR, PKS 1209-51/52 (G296.5+10.0), (Milne & Haynes 1994) with systematic RM variations.

## 4. Physical properties of the SNR G156.2+5.7

As shown by Gerardy & Fesen (2007), the analysis of multiband images reveals rich information of this SNR. To understand the physical properties of the SNR G156.2+5.7 and its environment better, we overlaid the λ6 cm radio map onto images obtained at other bands, such as at the broad-band ROSAT X-ray image, a low resolution Hα image, and the 100 μm IRAS image (Fig. 8). We rotated the composed image of the ROSAT X-ray image (blue), high resolution Hα image (green) and the $A_v$ extinction map (red) from Gerardy & Fesen (2007) and include it in this context for a direct comparison with the low-resolution radio images and the high-resolution Hα map.

From the high resolution Hα image (Gerardy & Fesen 2007), the Hα filaments observed on the left, the top and upper-right of Fig. 8 are reminiscent of the non-radiative shock emission along the outer edges of a SNR. Therefore, these filaments clearly define the X-ray boundary (blue). Very surprising is that no radio emission detected. In the shock region of a SNR one would expect a compressed ambient interstellar magnetic field with accelerated electrons and thus enhanced synchrotron emission. Furthermore, the very weak bulge of X-ray emission on the top is defined by the blowing-out of Hα filaments, while radio emission is completely absent at all bands at present sensitivity.

We noticed that Hα filaments and X-ray emission have a well defined spherical boundary in the upper-left part of the SNR, clearly indicating that the medium is much thinner and relatively uniform there. However, in the lower-right part, the boundary is not well defined and therefore the medium in that area is likely denser and more irregular. The interstellar medium around the SNR is rather inhomogeneous as pointed out by Gerardy & Fesen (2007). The extinction from foreground clouds is obviously larger in the lower part, and particular strong in the lower-right part, which is confirmed by the 100 μm IRAS image showing enhanced emission in this area. Therefore the X-ray emission in the lower part of the SNR has been significantly absorbed. Physical parameters estimated from X-ray spectrum of some parts of this remnant (Yamauchi et al. 1999; Pannuti & Allen 2004) should be treated with caution.

However, much stronger radio emission is detected in the lower-right part of the SNR having a similar open-shell structure and intensity enhancement in the radio total intensity and in the 100 μm emission as shown by the overlay of the corresponding images. This indicates a possible interaction between the SNR and molecular material. Obviously this SNR shows enhanced radio emission towards a denser interstellar environment.

Polarized radio emission has been detected from the SNR shell and a central patch. As shown in two left pannels in Fig. 8, the polarized left shell ridge is located inside the circular boundary of the remnant (defined by extrapolation from the upper-left Hα arc) but outside the bright Hα ridge. The magnetic field lines shown by our λ6 cm radio polarized emission nicely follow the extrapolated circular boundary. The right radio shell ridge coincides with the boundary defined by clumpy Hα emission. The magnetic field lines also follow the SNR shell boundary here. The central polarized radio patch is probably located in the front shell of the remnant, where enhanced radio emission results from interaction of the remnant with an interstellar cloud. It is also interesting to note that the magnetic field lines show a symmetric open-shell strcuture similar to the interstellar cloud as visible in the IRAS 100 μm image. This indicates a physical connection probably related to the SNR expansion or by stellar wind effects from the SNR progenitor.

## 5. Conclusions

We have observed the SNR G156.2+5.7 for the first time at λ6 cm wavelength and obtained a total intensity and linear polarization map with very good quality. We analysed the new radio measurements with other radio data from the literature and derived a spectral index map and a map of the RM distribution for this SNR. The RM gradients in opposite direction indicate the existence of a toroidal component in the SNR shell. The radio emission originates in the interaction region of the supernova shoch with dense interstellar clouds. The central polarized patch of G156.2+5.7 is a remarkable feature, which originates most likely in the front of the SNR shell. There is no indication for a central component inside the SNR.

*Acknowledgements.* The λ6 cm data were obtained with the receiver system from MPIfR mounted at the Nanshan 25-m telescope at the Urumqi Observatory of NAOC. We thank the staff of the Urumqi Observatory of NAOC for the great assistance during the installation of the receiver and the observations. In particular we like to thank Otmar Lochner for the construction of the 6 cm system and its in- stallation and M. Z. Chen and J. Ma for their help during the installa- tion of the receiver and its maintenance. We are very grateful to Peter Müller for software support. MPG and NAOC supported the construction of the Urumqi λ6 cm receiving system by special funds. The research work of JWX, LX, XHS and JLH was supported by the National Natural Science foundation of China (10473015 and 10521001) and by the Partner group of MPIfR at NAOC. JWX thanks C. Wang for his assistance and help during the data reduction process. We are grateful to E. Fürst and J. Dickel for constructive comments on the manuscript.